\documentclass{PoS}
\usepackage{amssymb,amsmath}

\newcommand{\ag}{{\alpha_G}}             % transplackian coupling
\newcommand{\ampRid}{\mathfrak{M}}       % reduced amplitude occurring in coherent state
\newcommand{\bk}[1]{\langle #1 \rangle}  % bra-ket
\newcommand{\bt}{{\boldsymbol{b}}}
\newcommand{\dif}{\mathrm{d}}            % finite-dimensional differential
\newcommand{\el}{{\mathrm{el}}}          % elastic
\newcommand{\esp}[1]{\mathrm{e}^{#1}}    % esponential
\newcommand{\haw}{H}%\mathrm{Hawking}}
\newcommand{\om}{\omega}
\newcommand{\tomE}{\textstyle{\frac{\om}{E}}}
\newcommand{\ord}[1]{\mathcal{O}\left(#1\right)}
\newcommand{\qt}{{\boldsymbol{q}}}
\newcommand{\tb}{\bar{\tau}}             % saddle point
\newcommand{\tfa}{{\cal M}}              % Fourier transform of amplitude
\newcommand{\ui}{\mathrm{i}}             % imaginary unit
\newcommand{\xt}{{\boldsymbol{x}}}

\title{Radiation enhancement and ``temperature'' in the collapse regime of gravitational scattering}

\ShortTitle{Radiation enhancement and ``temperature'' in the collapse regime of gravitational scattering}

\author{\speaker{Dimitri Colferai}\\
  (Dipartimento di Fisica, Universit\`a di Firenze and INFN Sezione di Firenze)\\
  E-mail: \email{colferai@fi.infn.it}}

\author{Marcello Ciafaloni\\
  (Dipartimento di Fisica, Universit\`a di Firenze)\\
  E-mail: \email{ciafaloni@fi.infn.it}}

\abstract{
  We generalize the semiclassical treatment of graviton radiation to
  gravitational scattering at very large energies $\sqrt{s}\gg m_P$ and finite
  scattering angles $\Theta_s$, so as to approach the collapse regime of impact
  parameters $b \simeq b_c \sim R\equiv 2G\sqrt{s}$. Our basic tool is the
  extension of the recently proposed, unified form of radiation to the ACV
  reduced-action model and to its resummed-eikonal exchange. By superimposing
  that radiation all-over eikonal scattering, we are able to derive the
  corresponding (unitary) coherent-state operator. The resulting graviton
  spectrum, tuned on the gravitational radius $R$, fully agrees with previous
  calculations for small angles $\Theta_s\ll 1$ but, for sizeable angles
  $\Theta_s(b)\leq \Theta_c = \ord{1}$ acquires an exponential cutoff of the
  large $\om R$ region, due to energy conservation, so as to emit a finite
  fraction of the total energy. In the approach-to-collapse regime of
  $b\to b_c^+$ we find a radiation enhancement due to large tidal forces, so
  that the whole energy is radiated off, with a large multiplicity
  $\bk{N}\sim Gs \gg 1$ and a well-defined frequency cutoff of order $R^{-1}$.
  The latter corresponds to the Hawking temperature for a black hole of mass
  notably smaller than $\sqrt{s}$.
}

\FullConference{The European Physical Society Conference on High Energy Physics\\
                 5-12 July\\
                 Venice, Italy}

\begin{document}

\section{Outline}

In this talk I show that a quantum description of particle scattering at extreme
energies can give rise to graviton radiation whose properties are similar to
Hawking radiation, thus suggesting a sort of gravitational collapse. However,
the radiation process can still be described by a unitary $S$-matrix, thus
pointing towards a resolution of the information paradox.

In order to explain the main ideas of our analysis, I shall give a brief
introduction to the ACV method for describing string or particle collisions at
transplanckian energies; then I'll describe our method for treating graviton
radiation which is based on a unified emission amplitude.  Finally I'll show and
discuss the main results: (i) the final state radiation produced in such
collisions is a unitary pure state; (ii) we see the role of the gravitational
radius in the graviton spectrum, since $\bk{\om} \sim 1/R$; (iii) the ensuing
energy spectrum presents a "quasi-temperature" $T$ of the order of the Hawking
temperature $T_\haw$.

\section{The ACV approach}

Following the pioneering work of Amati-Ciafaloni-Veneziano (ACV) \cite{ACV}, we
consider two strings colliding at center-of-mass energy $2E = \sqrt{s} \gg M_P$
much larger than the Planck mass, with some impact parameter $b$.  The condition
on the energy could lead, according to general relativity, to the formation of a
macroscopic black-hole whose radius $R = 2G \sqrt{s} \gg l_P,l_s$ is much larger
than both the Planck length and the string length. In this high-energy regime
the adimensional parameter $\alpha_G \equiv Gs/\hbar$ is very large and plays
the role of effective gravitational coupling. It measures also the typical
action of such processes, in units of $\hbar$.

ACV found that string scattering amplitudes at transplanckian energies and at
fixed momentum transfer (the so called Regge limit) can be interpreted as
effective ladder-like diagrams involving (Reggeized) graviton exchanges between
(excited) string states.

Since strings are extended objects, they interact by exchanging soft momenta, of
the order $\hbar/b$. On the other hand, they can exchange a large number of
quanta, on average $\langle n\rangle = Gs/\hbar$.  In this way the overall
exchanged momentum can be large: $Q\sim Gs/b$ yielding a finite deflection angle
$\Theta_E = Q/E \sim R/b$.

In the semiclassical regime $b\gtrsim R\gg l_P$, the string states are actually
not excited, i.e., on-shell, and can be treated as point particles.  The generic
ladder diagram is then just a convolution in transverse momentum of the single
rung amplitude.  This convolution can be diagonalized by Fourier transform from
transverse momentum $Q$ to impact parameter space $b$. The resummation of such
diagrams yields an exponential series, and provides the "eikonal" form of the
$S$-matrix for elastic scattering $S_\el=\esp{\ui2\delta(b,s)}$, where
$\delta(b,s)\equiv\ag\Delta(b)$ represents the phase shift.  From the latter,
one can compute some observables, e.g., the Einstein deflection angle
$\Theta_E = \sqrt{s}/(4\hbar)\partial\delta/\partial b = 2R/b$.

In order to approach the collapse region, we decrease the impact parameter
at values $b\sim R$. Here gravity is strong and subleading corrections to the
eikonal (of order $R^2/b^2$) become important. ACV were able to identify such
subleading diagrams~\cite{ACV2}, the so-called H diagrams.

The new ingredient of such diagrams is an effective vertex introduced by
Lipatov~\cite{Li91}.  By resumming also the subleading diagrams, ACV determined
the next-to-leading corrections to the phase shift and found a very interesting
feature~\cite{ACV07}: for impact parameters smaller than a critical value
$b_c\sim R$, the phase shift acquires a positive imaginary part, implying a
suppression of the elastic $S$-matrix. In particular, at vanishing impact
parameters, the elastic $S$-matrix is exponentially suppressed like
$|S_\el| \sim exp(-\pi Gs/\hbar)$.

A crucial question then arises: is such elastic suppression fully compensated by
inelastic graviton production which restores unitarity? Or is such critical
parameter the signal of something deeper, maybe the signal of a gravitational
collapse? Note that $b_c$ is a branch-cut singularity of $\Delta(b)$, with a
fractional critical exponent 3/2.

\section{Graviton radiation in transplanckian scattering}

The Lipatov vertex is appropriate only in the so-called multi-Regge kinematics,
where the emission angle of the graviton is much greater than the scattering
angle of the external particle.  In the complementary region of small emission
angles and relatively soft graviton energies ($\om \ll E$), amplitudes are well
described by the Weinberg current. Those two regions overlap, and we can write a
unified amplitude~\cite{CCCV15} describing in a simple and elegant way
graviton emission in both Regge region and soft region.
In particular, in coordinate space, where $b$ is the impact parameter and $x$
the transverse coordinate of the emitted graviton, such amplitude can be
expressed as the soft amplitude, minus the same soft amplitude with $E$ (the
energy of the external particle) replaced by $\omega$ (the energy of the emitted
graviton):
\begin{equation}\label{ampli}
  \tfa_\lambda = \sqrt{\ag}\frac{R}{2\pi^2}\esp{\ui\lambda\phi_\qt}
  \int\frac{\dif^2\xt}{|x|^2\esp{\ui\lambda\phi_x}} \;
  \esp{\ui\qt\cdot\xt}\left\{
    \frac{E}{\om}\left[\Delta(\bt)-\Delta\left(\bt-\tomE\right)\right]
    -\left[\Delta(\bt)-\Delta(\bt-\xt)\right] \right\} \;.
\end{equation}
We can interpret these two contributions as graviton insertions on
the external legs (\`a la Weinberg) and on the internal leg, respectively. We
refer to such decomposition as "soft-based" representation.

The single-graviton emission amplitude is then computed by resumming all ladder
diagrams with one emission.  When gravitons are emitted from within the ladder,
the local incoming particle momentum is rotated with respect to the $z$-axis.
Accordingly, amplitudes acquires phase factors. Denoting with $\theta$
($\theta_s$) the graviton emission angle (the scattering angle), if
$\theta_s\ll\theta$, such phase factors cancel and amplitudes interfere
constructively.  On the other hand, if $\theta\sim\theta_s$, this is no more the
case, and coherence is lost.  Quantitatively this has very important
consequences.

In addition, the emitted graviton can rescatter with the incident particles.
In particular, a graviton at transverse position $x$ mainly interacts with the
particle moving in the opposite direction at transverse position $b$, and the
interaction is given by the usual eikonal function, but now proportional to the
graviton energy omega and with shifted transverse distance $b-x$.

It turns out that the full $2\to 3$ amplitude can be written as the elastic
amplitude, times a factor $\ampRid$ which describes graviton production:
\begin{equation}
  \ampRid_\lambda(\bt;\om,\qt) =
  \sqrt{\ag}\frac{R\esp{\ui\lambda\phi_\qt}}{2\pi^2}
  \int\frac{\dif^2\xt}{|x|^2\esp{\ui\lambda\phi_x}}
  \frac{\esp{\ui\qt\cdot\xt}}{2\ui\om R}\left\{\esp{2\ui\om R\left[
        \Delta(\bt-\xt)-\Delta(\bt)\right]}-\esp{2\ui\om R\frac{E}{\om}\left[
        \Delta(\bt-\frac{\om}{E} \xt)-\Delta(\bt)\right]} \right\}
\end{equation}
The first term contains the effect of angular coherence discussed before, the
second term contains the rescattering effects.  It is important to note that the
same $\Delta$ function describes eikonal exchanges during the elastic scattering
and graviton production and rescattering!

The final step is to consider the emission of an arbitrary number of gravitons.
In the eikonal approximation such emissions are independent.  In fact, for 
$b\gg R$, we have shown that the amplitude for the emission of $N$ real
gravitons factorizes as a product of the elastic amplitude times an emission
factor $\ampRid_{\lambda_i}(q_i):i=1,\cdots,N$ for each graviton.
Such factorization is equivalent to assume that the $S$-matrix is given by the
elastic $S$-matrix times a coherent state operator acting on the Fock space of
gravitons: $S=\esp{\ui 2\delta}\exp\left\{\ui\sum_\lambda\int\frac{\dif^3 k}{2\omega_k}
    \ampRid_\lambda(k)a_\lambda^\dagger(k) + \text{h.c.}\right\}$.
The $a^\dagger$ term in the exponent is responsible for graviton
creation via the emission factor $\ampRid$, and we add the hermitian conjugate
term which effectively describes virtual corrections, according to Weinberg
analysis of soft graviton emission. In this way we have a unitary $S$-matrix
describing transplanckian collisions.

\begin{figure}
{\centering
\includegraphics[width=.4\textwidth]{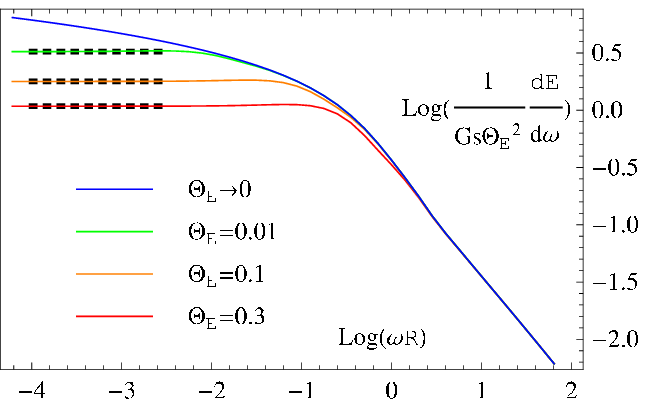}\hspace{0.1\textwidth}
\includegraphics[width=.4\textwidth]{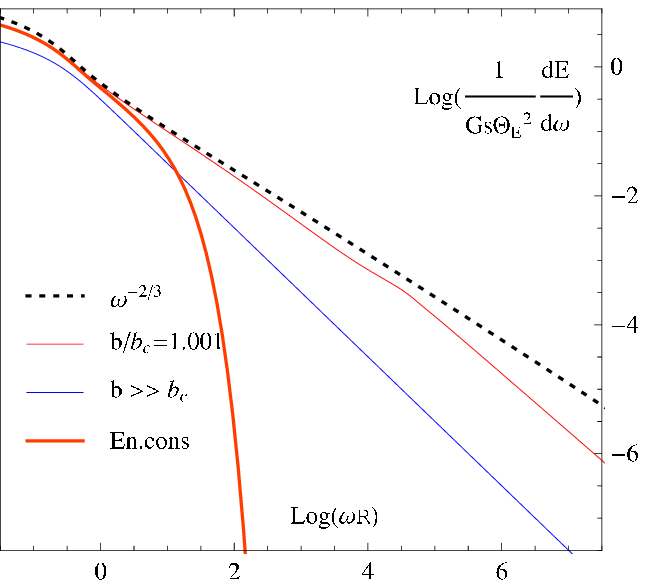}\\
}
\caption{Left (a): graviton energy spectrum for $b\gg R$, i.e., small deflection
  angles. Right (b): spectrum close to the critical parameter $b\simeq b_c$ with
  and without energy conservation.}
\label{f:spectrum}
\end{figure}

Using the previous result it is straightforward to compute the energy or
frequency ($\hbar=1$) spectrum of graviton radiation~\cite{CCCV15}. It is shown
in fig.~\ref{f:spectrum}a, for various values of the scattering angle, i.e., of
the ratio $R/b$. The overall radiated energy increases with $s$ and also with
the scattering angle.  However, the shape of the spectrum is weakly dependent of
such kinematical parameters: it is almost constant for small frequencies
$\om<1/R$, reproducing the expected zero-frequency limit, while it decreases
like $1/\om$ for $\om>1/R$.  The main feature of this spectrum is that is
clearly shows a characteristic frequency $\omega \sim 1/R$, almost independently
of the scattering angle.  This means that if we increase the energy of the
process, the frequency of the radiation decreases, in a way reminiscent of how
Hawking radiation depends upon the black-hole mass.

Now we want to move towards the strong coupling region $b\sim R$. Here, we have
to take into account (i) H diagrams along the ladder and (ii) graviton emissions
from H diagrams. The first effect is easily taken into account, by replacing the
leading eikonal phase $\Delta$ with the resummed one that develops an imaginary
part at $b = b_c$.  The second effect is more difficult to evaluate.  We
argue~\cite{CC16} that the sum of external and internal insertions of gravitons
into the H diagram can be again described by the "soft-based" representation
mentioned before.

By neglecting correlations between emitted gravitons, we end up with the same
formulae for amplitudes and spectra, where now the $\Delta$ function is the
resummed one. For $b$ approaching $b_c$, we note a significant enhancement of
radiation. In particular, there is a region where the spectrum decreases slowly,
like $\om^{-2/3}$, before eventually decreasing like $1/\om$
(fig.~\ref{f:spectrum}b). This is due to larger and larger tidal forces as $b$
decreases, and an increasing fraction of energy is radiated off.

\section{Energy conservation and ``temperature''}

At this point it is important to take into account at least those correlations
due to energy conservation.  We impose it event by event, by requiring
$\sum_i\omega_i < E$, and extend this bound to virtual corrections on the basis
of AGK cutting rules.
Without entering into technical details, we insert the step function
$\Theta(E-\sum_i\omega_i)$ in the individual emission probabilities. By using
the usual integral representation for the step function, we find that the
integral is dominated by a saddle point $\tb$, and the result~\cite{CC16} is
that the naive uncorrelated spectrum is suppressed by a factor $\esp{\om/\tb}$
(fig.~\ref{f:spectrum}b).  This behaviour is similar to thermal radiation, and
in our case the quasi-temperature approaches $T = 1/(\tb R) = 0.8/R \sim T_\haw$
at $b = b_c$, say the temperature of a black-hole of mass $\simeq 0.1 \sqrt{s}$.
But I want to stress that our radiation is coherent, it is a pure state!

Finally we want to cross the critical impact parameter and reach $b\ll R$, i.e.,
central collisions.  In this case the elastic amplitude provides an exponential
suppression $\exp(-\pi ER)$ of all graviton amplitudes.  However, the
rescattering term of the emission factor, which dominates in this regime, is
enhanced like $\exp(+\pi\om R)$. Therefore, if $\sum_j\om_j=E$, such
enhancement compensates the suppression, and there is the possibility of
recovering unitarity also in this subcritical regime.  Here analytic and
numerical calculations are more difficult, but a first estimate of the
quasi-temperature indicates a value $T \simeq 0.7/R$, similar to that found at
$b = b_c$ and again of the order of $T_\haw$.

In conclusion, we set up a framework which allows us to compute graviton
radiation in transplanckian collisions.
We now see the role of the gravitational radius as the inverse of the
characteristic frequency for all values of center-of-mass energies and impact
parameters.
When the impact parameter is $b\lesssim R$, tidal forces cause a dramatic
enhancement of graviton radiation and all energy is radiated off.  By taking
into account energy-conservation constraints, we obtain a coherent radiation
pattern with an exponentially decreasing spectrum, resembling a thermal
radiation with $T \sim T_\haw$ of a black-hole somewhat lighter than $\sqrt{s}$.
All in all, this radiation mechanism preserve unitarity and suggests a possible
solution of the information paradox.

\end{document}